\documentclass[letterpaper, 10 pt, conference]{ieeeconf}  

\IEEEoverridecommandlockouts                              
\usepackage[utf8]{inputenc}
\usepackage{amsthm}

\usepackage[noadjust]{cite}

\let\labelindent\relax
\usepackage{enumitem}

\usepackage{comment}

\usepackage{amsmath,amssymb, color}

\usepackage{tikz}
\usetikzlibrary{shapes,arrows,calc}
\definecolor{MyGreen}{RGB}{50,140,80}

\usepackage{ circuitikz }
\usetikzlibrary{arrows}

\title{\LARGE \bf
   Rhythmic neuromorphic control of a pendulum:  \\ A hybrid systems analysis
}

\author{E. Petri, R. Postoyan, W.P.M.H. Heemels 
\thanks{This work is partly funded by the ANR under grant OLYMPIA ANR-23-CE48-0006.}
\thanks{Elena Petri and  Maurice Heemels are with the Control Systems Technology (CST) Section of the Department of Mechanical Engineering,	Eindhoven University of Technology, The Netherlands
        (\{e.petri,m.heemels\}@tue.nl).}%
        \thanks{Romain Postoyan is with the Université de Lorraine,
CNRS, CRAN, 54000 Nancy, France 
         (romain.postoyan@univ-lorraine.fr).}}

\begin{document}

\newcommand\romain[1]{\emph{{\color{blue}#1}}}
\newcommand\elena[1]{\emph{{\color{orange}#1}}}
\newcommand\maurice[1]{\emph{{\color{green}#1}}}

\newcommand{\Rl}[2]{\ensuremath{\mathbb{R}^{#1}_{#2}}}   
\newcommand{\R}{\ensuremath{\mathbb{R}}}
\newcommand{\Rlp}{\ensuremath{\mathbb{R}_{>0}}}
\newcommand{\Rlpc}{\ensuremath{\overline{\mathbb{R}}_{>0}}}
\newcommand{\Rlo}{\ensuremath{\mathbb{R}_{\geq 0}}}
\newcommand{\Rln}{\ensuremath{\mathbb{R}_{< 0}}}
\newcommand{\Zo}{\ensuremath{\mathbb{Z}_{\geq 0}}}
\newcommand{\Zp}{\ensuremath{\mathbb{Z}_{> 0}}}
\newcommand{\N}{\ensuremath{\mathbb{N}}}                 
\newcommand{\Z}{\ensuremath{\mathbb{Z}}}

\definecolor{bleucit}{rgb}{0.2,0.4,0.6} 
\newcommand{\bleucit}{\textcolor{bleucit}}
\newcommand\cp[1]{{`\emph{#1}'}}
\newcommand\blue[1]{\emph{{\color{blue}#1}}}

\newcommand{\cmark}{\ding{51}}%
\newcommand{\xmark}{\ding{55}}%

\newcommand{\dst}{\displaystyle}
\newcommand{\Linf}[1]{\ensuremath{\mathcal{L}^{#1}}}

\newcommand{\eg}{{\it e.g.}}

\newcommand{\Nesic}{Ne{\v{s}}i{\'c} }

\definecolor{blue_cv}{rgb}{0.09,0.35,0.78}

\newcommand{\KL}{\ensuremath{\mathcal{KL}}}
\newcommand{\K}{\ensuremath{\mathcal{K}}}
\newcommand{\Kinf}{\ensuremath{\mathcal{K}_{\infty}}}
\newcommand{\KK}{\ensuremath{\mathcal{KK}}}
\newcommand{\KN}{\ensuremath{\mathcal{KN}}}
\newcommand{\KKL}{\ensuremath{\mathcal{KKL}}}
\newcommand{\KLL}{\ensuremath{\mathcal{KLL}}}
\newcommand{\D}{\ensuremath{\mathcal{D}}}
\newcommand{\PD}{\ensuremath{\mathcal{PD}}}

\newcommand{\Cs}{\ensuremath{C_{\text{steady}}}}
\newcommand{\Ct}{\ensuremath{C_{\text{transient}}}}
\newcommand{\Ds}{\ensuremath{D_{\text{steady}}}}
\newcommand{\Dt}{\ensuremath{D_{\text{transient}}}}
\newcommand{\UtGpAS}{U$_{\text{t}}$GpAS}
\newcommand{\UtGAS}{U$_{\text{t}}$GAS}
\newcommand{\UjGpAS}{U$_{\text{j}}$GpAS}
\newcommand{\UjGAS}{U$_{\text{j}}$GAS}

\newcommand{\argmin}{\ensuremath{\text{argmin}\,}}
\newcommand{\interior}{\ensuremath{\text{int}\,}}
\newcommand{\dom}{\ensuremath{\text{dom}\,}}
\newcommand{\Span}{\ensuremath{\text{Span}}}
\newcommand{\avg}{\ensuremath{\text{avg}}}
\newcommand{\co}{\ensuremath{\text{co}\,}}
\newcommand{\coc}{\ensuremath{\overline{\text{dom}}\,}}
\newcommand{\ext}{\ensuremath{\text{ext}}}
\newcommand{\rge}{\ensuremath{\text{rge}\,}}
\newcommand{\esup}{\ensuremath{\text{ess.sup}\,}}
\newcommand{\sign}{\ensuremath{\text{sign}}}
\newcommand{\SGN}{\ensuremath{\text{SGN}}}

\newcommand{\sat}{\ensuremath{\text{sat}}}

\newcommand{\sinc}{\ensuremath{\text{sinc}}}
\newcommand{\nom}{\ensuremath{\text{nom}}}

\newcommand{\Tmati}{\ensuremath{T_{MATI}\,\,}}
\newcommand{\Tmasp}{\ensuremath{\mathrm{T_{MASP}}\,}}
\newcommand{\lc}{\ensuremath{\llbracket}}
\newcommand{\rc}{\ensuremath{\rrbracket}}

\newcommand{\norm}[1]{\ensuremath{\left\|{#1}\right\|}}
\newcommand{\ip}[2]{\ensuremath{\left\langle #1, #2\right\rangle}}
\newcommand{\cb}[1]{\ensuremath{\overline{\mathbb{B}}_{\mbox{\scriptsize $#1$}}}}                              
\newcommand{\ob}[2]{\ensuremath{\mathbb{B}_{\mbox{\scriptsize $#1$}}\ensuremath{\left( #2\right)}}}   
\newcommand{\df}{\ensuremath{\stackrel{\mbox{\tiny $\mathrm{def}$}}{=}\:}}                                             
\newcommand{\myint}[4]{\ensuremath{\int_{#1}^{#2}#3\;\mathrm{d}#4}}
\newcommand{\Mm}{\ensuremath{\:\stackrel{\rightarrow}{\scriptstyle{\rightarrow}}\:}}
\newcommand{\bm}[1]{\ensuremath{\mathbf{#1}}}
\newcommand{\hs}[1]{\hspace*{#1 em}}
\newcommand{\qa}{\ensuremath{\mathcal{Q}_{A}}}%
\newcommand{\mc}[1]{\ensuremath{\mathcal{#1}}}
\newcommand{\di}{\ensuremath{\mathcal{D}_{i}}}

\newcommand{\HS}{\ensuremath{\mathcal{H}}}
\newcommand{\HSc}{\ensuremath{\mathcal{H}_c}}


%


\newtheorem{exple}{Example}
\newtheorem{defn}{Definition}
\newtheorem{claim}{Claim}
\newtheorem{hypo}{Hypothesis}
\newtheorem{ass}{\textnormal{\textbf{Assumption}}}
\newtheorem{prop}{Proposition}
\newtheorem{fact}{Fact}
\newtheorem{lem}{Lemma}
\newtheorem{ex}{Example}
\newtheorem{thm}{\textnormal{\textbf{Theorem}}}
\newtheorem{cor}{\textnormal{\textbf{Corollary}}}
\newtheorem{pb}{Problem}
\newtheorem{rem}{\textnormal{\textbf{Remark}}}
\newtheorem{sass}{Standing Assumption}


%
%
\newenvironment{rems}{\textit{Remarks. }}{\mbox{}\\[1ex]}

\maketitle
\thispagestyle{empty}
\pagestyle{empty}

\newcommand{\hide}[1]{}
\begin{abstract}
Neuromorphic engineering is an emerging research domain that aims to realize important implementation advantages that brain-inspired technologies can offer over classical digital technologies, including energy efficiency, adaptability, and robustness. For the field of systems and control, neuromorphic controllers could potentially bring many benefits, but their advancement is hampered by lack of systematic analysis and design tools. In this paper, the objective is to show that hybrid systems methods can aid in filling this gap.  We do this by formally analyzing rhythmic neuromorphic control of a  pendulum system, which was recently proposed as a prototypical setup. The neuromorphic controller generates spikes, which we model as a Dirac delta pulse, whenever the pendulum angular position crosses its resting position, with the goal of inducing a stable limit cycle. This leads to modeling the closed-loop system  as a hybrid dynamical system, which in between spikes evolves in open loop and where the jumps correspond to the spiking control actions.  Exploiting the hybrid system model, we formally prove the existence, uniqueness, and a stability property of the hybrid limit cycle for the closed-loop system. Numerical simulations illustrate our  approach. We finally elaborate on a possible spiking adaptation mechanism on the pulse amplitude to generate a hybrid limit cycle of a desired maximal angular amplitude.

\end{abstract}

\section{Introduction}

Inspired by brain computational principles, neuromorphic engineering consists of developing hardware and software that emulate biological neural systems \cite{mead1990neuromorphic}. These advantages have motivated numerous research activities in various engineering fields in the recent years, see, e.g., \cite{gallego2022event, krauhausen2021organic, shrestha2022survey, yamazaki2022spiking, singh2018regulation}  and the references therein. 
This field holds significant promise for control over classical digital technologies in terms of energy efficiency, real-time processing, adaptive learning, and robust integration of sensory actuators \cite{ribar2021neuromorphic,mead1990neuromorphic}, but until now very little has been done in the context of systems and control, although inspiring exceptions exist, see, e.g., \cite{deweerth1990neuron,deweerth1991simple,sepulchre2022spiking,schmetterling2024neuromorphic, feketa2023artificial}. It is fair to say that neuromorphic control is still in its early development, for which one of the main causes is the lack of methodological tools for its design and analysis. This paper aims to contribute to filling this gap by providing systematic analysis and design tools for neuromorphic control systems.   

Due to the event-based nature of biological neurons, see, e.g., \cite{gerstner2002spiking,sepulchre2022spiking}, caused by the fact that information is communicated via the occurrence and the absence of spikes, we argue that hybrid systems theoretical tools as in, e.g., \cite{goebel2012hybrid,liberzon2003switching,brogliato2003some} are well suited to analyze and design neuromorphic controllers. To show this, following the preliminary path laid down in \cite{petri2024analysis}, where a stabilizing neuromorphic control is discussed for a simple scalar-state linear system, the aim of the present work is to show how the hybrid dynamical systems formalism in \cite{goebel2012hybrid} can be exploited to study more complex dynamical scenarios in neuromorphic control.  In particular, we formally investigate the scenario presented in \cite{schmetterling2024neuromorphic} in which a mechanical pendulum system is sporadically controlled by bursting neurons to generate desirable oscillations. 
The idea in \cite{schmetterling2024neuromorphic} is to activate a torque for a fixed  amount of time using bursting neurons whenever the pendulum  is at its resting position. In doing so, oscillations are generated \cite{schmetterling2024neuromorphic}, whose amplitude can be controlled by adapting the torque activation length through neuron parameters modifications, similar to neuromodulation in nature. This setting was deliberately chosen in \cite{schmetterling2024neuromorphic} as it can be seen as a prototypical setup that demonstrates the idea of entraining a mechanical system with a neuro-inspired controller that connects the central pattern generators
of neuroscience and the rhythmic controllers of legged robotics
\cite{ijspeert2008central,angelidis2021spiking}.  The objective of this work is to provide an analytical justification of the results presented in \cite{schmetterling2024neuromorphic} using hybrid systems tools \cite{goebel2012hybrid,lou-et-al-tac18(hybrid-limit-cycle)}, thus underlining the potential of hybrid systems for providing formal analysis and development tools for neuromorphic controllers.



To study the pendulum setup in \cite{schmetterling2024neuromorphic}, we model  the activation of bursting neurons as a Dirac delta pulse on the angular velocity (Section \ref{sect:ProblemStatement}). In this way, the duration of the activation is related to the amplitude of the pulse. After linearizing the pendulum dynamics, we present a hybrid model of the closed-loop system for pulses of a fixed  amplitude (Section \ref{sect:hybrid-model}). We then establish the existence of a unique hybrid limit cycle \cite{lou-et-al-tac18(hybrid-limit-cycle)}, and we prove that this hybrid limit cycle exhibits uniform exponential stability properties (Section \ref{sect:existence-stability-limit-cycles}). We exploit for this purpose the features of the problem, in particular expressions of the solutions to the hybrid system. The guaranteed stability properties inherit robustness guarantees from \cite[Chapter 7]{goebel2012hybrid} and \cite{cai2009characterizations} due to the regularity of the problem, the compactness of the hybrid limit cycle and its stability properties. These results are numerically illustrated (Section \ref{sect:numerical-illustration}). 
We finally elaborate on the possibility of adapting the pulse amplitude to generate a hybrid limit cycle of a desired maximum angular amplitude (Section \ref{sect:discussion}).  
%
Finally,  conclusions and directions for future work are discussed (Section \ref{sect:conclusion}). Interestingly, our work is developed in parallel with the complementary paper \cite{medvedeva-et-al-cdc25}  also studying the same pendulum problem but using alternative methodological tools; see also the Acknowledgements section at the end of the paper. More precisely, the paper \cite{medvedeva-et-al-cdc25} uses a describing function analysis to study limit cycles, and designs an adaption unit for the pulse amplitude to obtain a desirable limit cycle of described maximal amplitude by exploiting a robust optimization problem. Both papers provide complementary insights and results that are of independent interest. 

\section{Preliminaries}\label{Notation}

Let $\R$ be the set of real numbers, $\Rlo$, $\Rlp$ and $\R_{<0}$  be the set of non-negative, strictly positive, strictly negative  real numbers, respectively. Let $\Z$ be the set of integers, $\Zo$ and $\Zp$ be the set of non-negative and strictly positive integers, respectively. The empty set is denoted $\emptyset$. Given any $x\in\Rl{n}{}$ and $y\in\Rl{m}{}$ with $n,m\in\Zp$, the notation $(x,y)$ stands for $[x^{\top},\,y^{\top}]^{\top}$. Given $x\in\R^{n}$ and a non-empty set $\mathcal{A}\subset\R^{n}$ with $n\in\Zp$, $|x|$ stands for the Euclidean norm of $x$ and $|x|_{\mathcal{A}}$ for the distance of $x$ to $\cal A$, i.e., $|x|_{\mathcal{A}}=\inf\{|x-z|\,:\,z\in\mathcal{A}\}$. We denote the identity matrix $I_{\text{d}}$, whose dimensions depend on the context. 
The Dirac delta function is denoted $\delta$. A function $\alpha:\Rlo\to\Rlo$ is of class $\Kinf$ if it is continuous, zero at zero, strictly increasing and unbounded. We use $\sign$ for the mapping from $\R$ to $\{-1,0,1\}$ such that $\sign(z)=1$ if $z>0$, $\sign(0)=0$ and $\sign(z)=-1$ if $z<0$. We also consider the set-valued map $\SGN:\R\rightrightarrows\{-1,1\}$  defined as, for any $z\in\R$, $\SGN(z)=\{1\}$ if $z>0$, $\SGN(z)=\{-1\}$ if $z<0$, and $\SGN(0)=\{-1,1\}$. 
For a set-valued map $F:\R^{n}\rightrightarrows\R^{m}$ with $n,m\in\Zp$, $\dom F:=\{x\in\R^n\,:\,F(x)\neq \emptyset\}$ and $\text{rge}\, F:=\{y\in\R^m\,:\,\exists x\in\R^n,\,\, y\in F(x)\}$; these concepts are similarly defined for single-valued maps. The tangent cone to the set $S\subset \R^{n}$ with $n\in\Zo$ at the point $x\in\R^n$ is denoted $\mathcal{T}_{S}(x)$, see \cite[Definition 5.12]{goebel2012hybrid}. 

Based on the formalism of \cite{goebel2012hybrid}, we will model the proposed closed-loop system as a hybrid system of the form
\begin{equation}
	 \left\{
	\begin{array}{rcll}
		\dot x &=& f(x) & \quad x \in \mathcal{C}
		\\
		x^+ & \in & G(x)  &\quad x \in \mathcal{D},
	\end{array}
	\right.
	\label{eq:hybridSystemNotation}
\end{equation}
where 
$\mathcal{C}\subseteq \R^{n_x} $ is the flow set, 
$\mathcal{D}\subseteq \R^{n_x}$ is the jump set,
$f:\R^{n_x}\to\R^{n_x}$ is the flow map and $G:\R^{n_x}\rightrightarrows\R^{n_x}$ is the jump (set-valued) map. 
We consider \emph{hybrid time domains} as in \cite[Definition 2.3]{goebel2012hybrid}.  The notation $(t,j) \geq (t^\star, j^{\star})$ means that $t \geq t^\star$ and $j \geq  j^{\star}$,  where $(t,j),(t^\star,j^\star)\in\R_{\geq 0}\times\Zo$.
We use the notion of solutions for system \eqref{eq:hybridSystemNotation}  in \cite[Definition~2.6]{goebel2012hybrid}. A solution $x$ is \emph{maximal} when it cannot be extended \cite[Definition 2.7]{goebel2012hybrid} and  is \emph{complete} when $\dom x$ is unbounded. We say that system \eqref{eq:hybridSystemNotation} satisfies the \emph{hybrid basic conditions} \cite[Assumption 6.5]{goebel2012hybrid} if: \emph{(i)} $f$ is continuous and $C\subset\dom f$; \emph{(ii)} $G$ is outer semicontinuous and locally bounded relative to $\mathcal{D}$ and $\mathcal{D}\subset\dom G$; \emph{(iii)} $\cal C$ and $\cal D$ are closed. 

\section{Problem statement}\label{sect:ProblemStatement} 
We consider the dimensionless model of a pendulum  given by 
\begin{equation}
	\ddot{q} + \alpha \dot{q} + \sin q = u,
	\label{eq:pendulum}
\end{equation}
where $q \in [-\pi, \pi]$ is the angular position of the pendulum, with $q = 0$ the resting position,  $\alpha \in (0,2)$ is the dimensionless damping and $u \in \R$ is the dimensionless torque use to control system \eqref{eq:pendulum}. 

Inspired by \cite{schmetterling2024neuromorphic}, a spiking torque, generated by a neuromorphic device, is applied whenever $q$ is at its resting position ($q = 0$) to make the pendulum swing. We denote by $\{t_i\}_{i \in \mathcal{Z}}$ with $\mathcal{Z}=\{1,\ldots, N\}\subset\Zp$, where $N\in \Zp$ or $N=\infty$, the sequence of spiking instants, defined as, for any $i\in\mathcal{Z}$,
\begin{equation}
	t_{i +1}:= \inf\{t> t_{i}: q(t_i) = 0\} \quad \text{with} \quad t_{0} =0,
	\label{eq:triggeringRuleSpikingTimesGlobal}
\end{equation}
with $t_{i+1}=\infty$ if $i+1\notin\mathcal{Z}$. The spiking torque can be written as, for any $t\in [t_i,t_{i+1})$ and $i\in\mathcal{Z}$,
	\begin{equation}
		u(t) = \sum_{k = 1}^{i}I\delta(t-t_k) w(t) \quad \text{with} \quad w(t)\in \text{SGN}(\dot q(t)),
		\label{eq:spikingInput}
	\end{equation} 
where $I \in \R_{> 0}$ is a design parameter. We will elaborate on the scenario where the value of $I$ is adapted to generate oscillations of desired maximal angular amplitude in Section~\ref{sect:discussion}, see also \cite{medvedeva-et-al-cdc25} for a more in-depth treatment. 
This torque can be approximately generated using bursting neurons, producing $I$ spikes of unitary amplitude \cite{schmetterling2024neuromorphic}. With this spiking torque, we aim to generate a desirable limit cycle as experimentally done in \cite{schmetterling2024neuromorphic}. 

The goal of this work is to analyse the existence and the stability properties of limit cycles for the closed-loop system \eqref{eq:pendulum}, \eqref{eq:spikingInput}. We propose for this purpose to model 
the  corresponding closed-loop system as a hybrid dynamical system as in (\ref{eq:hybridSystemNotation}), for which a jump corresponds to the spiking control action $u$. We then establish the existence of so-called hybrid limit cycles as formalized later which we prove to exhibit stability properties.

\section{Hybrid model of closed-loop system}\label{sect:hybrid-model} 

\hide{
\textcolor{red}{Note that we have not included (yet) the virtual angular jump (\emph{angular flip}) to constrain $q$ to take values in $[-\pi,\pi]$ in the hybrid model. This should be possible using similar lines as in Romain's draft. However, if we consider a linearized version of the pendulum, the model is valid only near $q_1 =0$ and we can prove local properties. In this case it would probably make more sense to have some constraints on the initial conditions and, probably also an upperbound on $u$ which guarantees that the angular flip cannot occur and we do not go beyond the constraints set.  }}

In this section, we first introduce the variables needed to derive the desired hybrid model. We also linearize the dynamics in (\ref{eq:pendulum}) around the origin to facilitate computations in the sequel. We then formalize the objective using hybrid systems terminology. 

\subsection{Variables}\label{subsect:variables}
\hide{\elena{At the moment we do not have conditions to guarantee $q_1 \in [-\pi,\pi]$. We cannot add the angular flip in \eqref{eq:hybrid-model-compact}, because in that case it will not satisfy the conditions in \cite[Theorem 1]{lou-et-al-tac18(hybrid-limit-cycle)}. I will try to think and find the conditions/ bound on initial condition and $I$ so that we ensure $q_1 \in [-\pi,\pi]$.}}

We denote by $q_1\in[-\pi,\pi]$ the pendulum angle and $q_2\in\R$ the angular velocity for the sake of convenience. We thus derive from \eqref{eq:pendulum} that between two consecutive jumps
\begin{equation}   
	\begin{array}{rlllll}
		\dot q_1 & = & q_2 \\
		\dot q_2 & = & -\alpha q_2 - \sin(q_1). 
	\end{array}
	\label{eq:sys-pendulum-q1-q2-flow}
\end{equation}
For ease of computations,  we consider a linearized version of \eqref{eq:sys-pendulum-q1-q2-flow} given by
\begin{equation}   
	\begin{array}{rlllll}
		\dot q_1 & = & q_2 \\
		\dot q_2 & = & -\alpha q_2 - q_1, 
	\end{array}
	\label{eq:sys-pendulum-q1-q2-flow_linearized}
\end{equation}
where we recall that $\alpha \in (0,2)$. The linearization is performed for $q_1=0$, we will thus be concentrating on the behaviour of the pendulum for ``moderate'' angular values. Hence, for the sake of convenience we will be considering $q_1\in\R$ in the sequel thereby ignoring possible virtual  ``angular flips'' when $q_1\in\{-\pi,\pi\}$;  otherwise a hybrid mechanism can very well be added to the hybrid model presented below. 

\begin{figure}
	\begin{center}
		\begin{tikzpicture}[scale=0.75,transform shape] 
			\fill[black] (0,0) circle (2pt); 
			\draw[dashed] (0,0) -- (0,-3);
			\draw[thick] (0,0) -- (2,-2);
			\fill[black] (2,-2) circle (6pt); 
			\draw[thick,->] (0,-1) arc[start angle=-90, end angle=-30, radius=0.7];
			\node at (0.5,-1.2) {\(q_1\)};
			\node[below] at (0,-3.2) {\small{Resting position $(q_1=0)$}};
			\node[below] at (1,-2.7) {\small{$\sigma=1$}};
			\node[below] at (-1,-2.7) {\small{$\sigma=-1$}};
		\end{tikzpicture}
        \vspace{-0.2cm}
		\caption{Illustration of the pendulum.}\label{fig:pendulum}
	\end{center}
    \vspace{-0.6cm}
\end{figure}
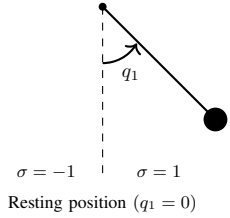

When $q_1$ crosses the resting position, a jump occurs and the spiking torque is applied to the pendulum. Thus, from \eqref{eq:spikingInput}, we have when $q_1=0$
\begin{equation}  
	\begin{array}{rlllll} 
		q_1^+ & = & 0 \\
		q_2^+ & \in & \{q_2 + Iz\,:\,z\in\SGN{(q_2)}\}. 
	\end{array}
	\label{eq:sys-pendulum-q1-q2-jump}
\end{equation}
Finally, we introduce an auxiliary logic variable $\sigma\in\{-1,1\}$ to robustify the detection of the condition $q_1=0$ needed to activate the torque. On flows,
\begin{equation}
	\dot\sigma = 0,
	\label{eq:sigmaFlow}
\end{equation}
and at jumps
\begin{equation}
	\sigma^+ \in \SGN{(q_2)}. 
	\label{eq:sigmaJump}
\end{equation}
Let us give some insights regarding the choice of $\sigma$. We have seen that the spiking torque must occur whenever $q_1=0$ on the one hand, and flow should be allowed for any $q_1\neq 0$. We may be tempted to define the flow set as $\{(q_1,q_2)\,:\,q_1\neq 0\}$ and the jump set as $\{(q_1,q_2)\,:\,q_1=0\}$. However, such a reasoning leads to a flow set, which is not closed thereby violating one of the hybrid basic conditions, namely \emph{(iii)} in Section \ref{Notation}. Variable $\sigma$ allows overcoming this shortcoming. The idea is to enforce $\sigma$ to have the same sign as $q_1$, as depicted in Fig. \ref{fig:pendulum}. In this way,  flow is only allowed when $\sigma q_1\geq 0$ and the spike occurs only when $\sigma q_1=0$ and $\sigma q_2\leq 0$. As a result, we will see next that the obtained flow and jump sets enforce the spiking torque activation as desired when $q_1=0$, while working with closed flow and jump sets thereby satisfying one of the hybrid basic conditions.

\subsection{Hybrid model}\label{subsect:hybridModel}

Given $I\in\Rlp$, we denote the concatenated state variable 
\begin{equation}
	\begin{array}{rlllll} 
		x & := & (q_1,q_2,\sigma)\in \mathbb{X}\subset\R^{n_x}\\
		\mathbb{X} & := & \R\times\R\times\{-1,1\},
	\end{array}
\end{equation}
with $n_x:=3$. We propose the hybrid model
\begin{equation}\label{eq:hybrid-model-compact}
	\begin{array}{rlllll} 
		\mathcal{H}_{I}\,:\,\left\{\begin{array}{rlllll}\dot x & = & f(x)  & & x\in\mathcal{C}\\
			x^+ & \in & G(x)   & & x\in\mathcal{D},
		\end{array}\right.
	\end{array}
\end{equation}
with 
\begin{equation}\label{eq:def-C-D}
	\begin{array}{rlllll} 
		\mathcal{C} & := & \left\{x\in\mathbb{X}\,:\,\sigma q_1\geq 0\right\} \\
		\mathcal{D} & := & \left\{x\in\mathbb{X}\,:\,\sigma q_1= 0,\,\sigma q_2\leq 0\right\} 
	\end{array}
\end{equation}
and
\begin{equation}\label{eq:def-f-G}
	\begin{array}{rlllll} 
		f(x) & := & (q_2,- q_1 -\alpha q_2,0) & x\in\mathcal{C}\\
		G(x) & := & \{(0,q_2 + Iz,z)\,:\,z\in\SGN{(q_2)}\}  &  x\in \mathcal{D}. \\
	\end{array}
\end{equation}
As explained at the end of Section \ref{subsect:variables}, flow is only allowed when $\sigma q_1\geq 0$. When $\sigma q_1=0$, a spiking torque occurs if in addition $\sigma q_2\leq 0$, see the definition of set $\mathcal{D}$. In this case, a jump of amplitude $I$ occurs on the angular velocity, which is positive or negative depending on the sign of $q_2$. 

\begin{rem} It can be noticed that when a solution to $\mathcal{H}_I$ is initialized in $\{0\}\times\{0\}\times\{-1,1\}$, it may remain there for all future times without ever experiencing a jump and thus not converge to a limit cycle. This case will be ruled out in the sequel by focusing on solutions not initialized in  $\{0\}\times\{0\}\times\{-1,1\}$, see Section \ref{subsect:set-X0}. 
\end{rem}

System $\mathcal{H}_I$ satisfies the next properties. 

\begin{prop}\label{prop:hybrid-model-basic-properties} Given any $I\in\R_{>0}$, the following holds for system $\mathcal{H}_I$ in (\ref{eq:hybrid-model-compact}).
\begin{enumerate}
\item[(i)] The hybrid basic conditions are satisfied.
\item[(ii)] Maximal solutions are complete.
\end{enumerate}
\end{prop}

\noindent\textbf{Proof:} Let $I\in\Rlp$. \emph{(i)} Flow and jump sets $\mathcal{C}$ and $\mathcal{D}$ in (\ref{eq:def-C-D}) are closed subsets of $\R^{n_{x}}$. In addition, $f$ is continuous and and $G$  is outer semicontinuous (its graph being closed) and locally bounded relative to $\mathcal{D}$ and $\dom G = \mathcal{D}$.  Hence \cite[Assumption 6.5]{goebel2012hybrid} is verified.

\noindent \emph{(ii)} We prove the desired result by verifying that the conditions of \cite[Proposition~6.10]{goebel2012hybrid} hold. Let $x=(q_1,q_2,\sigma)\in\mathcal{C}$. If $x$ is in the interior of $\mathcal{C}$, i.e., $\sigma q_1>0$, the tangent cone to $\mathcal{C}$ is given by $\mathcal{T_{C}} (x) = \R\times\R\times\{0\}$. If $x$ is on the boundary of $\mathcal{C}$, i.e., $\sigma q_1=0$, $\mathcal{T_{C}} (x) = \sigma\R\times\R\times\{0\}$. In view of the definition of $f$ in (\ref{eq:def-f-G}) and of $\mathcal{D}$, we observe that, for any  $x\in\mathcal{C}\backslash\mathcal{D}$, $f(x)\in\mathcal{T}_{\mathcal{C}}(x)$. Moreover, it can also be seen that for each $x\in \mathcal{C}\backslash\mathcal{D}$, there exists a neighborhood $\mathcal{U}$ of $x$ such that $\mathcal{U}\cap \mathcal{C}\cap \mathcal{D}=\emptyset$. Given the last two observations,  we deduce that \cite[(VC)]{goebel2012hybrid} holds. On the other hand, \cite[Proposition~6.10(b)]{goebel2012hybrid} cannot occur as $f$ is linear. Finally, $G({\mathcal{D}})\subset\mathcal{C}$ so that \cite[Proposition 6.10(c)]{goebel2012hybrid} cannot occur. We conclude that  Proposition~\ref{prop:hybrid-model-basic-properties}\emph{(ii)} holds by application of \cite[Proposition~6.10]{goebel2012hybrid}. \hfill $\blacksquare$

\subsection{Hybrid limit cycles}

Our objective is to establish the existence and the stability of hybrid limit cycles for system $\mathcal{H}_I$ for any given $I\in\R_{>0}$ as formalized next using the terminology introduced in \cite{lou-et-al-tac18(hybrid-limit-cycle)}, which we revisit and adapt to our purpose.

\begin{defn}\label{def:periodic-solution-hybrid-limit-cycle} Given any $I\in\Rlp$, a complete solution $x^\star$ to $\mathcal{H}_I$ is \emph{$(T^\star,N^\star)$-periodic} if $(T^\star,N^\star)\in\Rlo\times\Zo$ is such that the following holds.
\begin{enumerate}
\item[(i)] $T^{\star}+N^\star> 0$.
\item[(ii)] For all $(t,j)\in\dom x^\star$, $(t+T^\star,j+N^\star)\in\dom x^\star$.
\item[(iii)] For all $(t,j)\in\dom x^\star$ $x^\star(t+T^\star,j+N^\star)=x^\star(t,j)$.
\end{enumerate}
We call  $\mathcal{O}_I:=\text{rge}\,x^\star$ a \emph{hybrid limit cycle}. 
\end{defn}
The next definition formalizes the stability property we are interested in for the hybrid limit cycle generated by system $\mathcal{H}_I$. 

\begin{defn}\label{def:stability-hybrid-limit-cycle} Given any $I\in\Rlp$, consider system $\mathcal{H}_I$ and a set $\mathbb{X}_0\subset\mathcal{C}\cup\mathcal{D}$. We say that the hybrid limit cycle $\mathcal{O}_I$ is \emph{uniformly exponentially stable on $\mathbb{X}_0$} when the following properties hold.
\begin{enumerate}
\item[(i)] $\mathbb{X}_0$ is forward invariant, i.e., $\text{rge}(x)\subset\mathbb{X}_0$ for any solution $x$ initialized in $\mathbb{X}_0$.
\item[(ii)] There exist\footnote{Function  $\alpha$ may not be linear, as  customary when referring to exponential stability in the literature. As the decaying term is exponential, we argue that the stated stability property can still be qualified as being uniform and exponential.} $\alpha\in\Kinf$ and $\gamma\in\Rlp$ such that any solution $x$ initialized in $\mathbb{X}_0$  verifies $|x(t,j)|_{\mathcal{O}_I}\leq  e^{-\gamma(t+j)}\alpha(|x(0,0)|_{\mathcal{O}_I})$ for all $(t,j)\in\dom x$. 
\end{enumerate}
\end{defn}



\section{Existence and stability of a hybrid limit cycle}\label{sect:existence-stability-limit-cycles}

Unfortunately, we cannot invoke the existing results of the literature on the existence and the stability of limit cycles for hybrid systems of the form (\ref{eq:hybridSystemNotation}) to prove the desired properties for system $\mathcal{H}_I$ given $I\in\R_{>0}$, see \cite{lou-et-al-tac18(hybrid-limit-cycle),lou2023notions}. Indeed, as we will see in this section, the periodic solutions do not exhibit the right number of jumps over one ``hybrid period''  required by\footnote{By (\ref{eq:def-C-D}), taking $h(x)=\sigma q_1$, we have $\mathcal{C}=\{x\in\mathbb{X}\,:\,h(x)\geq 0\}$ and $\mathcal{C}=\{x\in\mathbb{X}\,:\,h(x)= 0,\,L_f h(x)\leq 0\}$ with the notation of \cite{lou-et-al-tac18(hybrid-limit-cycle),lou2023notions}. In this case, these references require $N^\star=1$, while we will see in Theorem~\ref{Thm:ExistenceFlowPeriodicSolution} that $N^{\star}=2$.} \cite{lou-et-al-tac18(hybrid-limit-cycle),lou2023notions} for system $\mathcal{H}_I$. Moreover, we aim at concluding stronger properties, namely an exponential stability property as in Definition \ref{def:stability-hybrid-limit-cycle}. We provide instead an analysis tailored to system $\mathcal{H}_I$. We first determine the set $\mathbb{X}_0$ and prove it is forward invariant as required by  Definition~\ref{def:periodic-solution-hybrid-limit-cycle}\emph{(i)} (Section \ref{subsect:set-X0}). Then, we state  insightful  expressions of the solution to $\mathcal{H}_I$ (Section \ref{subsect:explicitSolution}), which allow deriving properties on the inter-event times of the solutions initialized in $\mathbb{X}_0$. Given these results, we can prove that  there exists a unique hybrid limit cycle  in the sense of Definition~\ref{def:periodic-solution-hybrid-limit-cycle} (Section \ref{subsect:existence-limit-cycle}). We finally establish the desired stability property for this hybrid limit cycle (Section~\ref{subsec:StabilityHybridLimitCycle}).

\subsection{Forward invariant set $\mathbb{X}_0$}\label{subsect:set-X0}

We define the set $\mathbb{X}_0$ as
\begin{equation}\label{eq:X0}
\mathbb{X}_0 := \left(\mathcal{C}\cup\mathcal{D}\right)\backslash\left(\{0\}\times\{0\}\times\{-1,1\}\right).
\end{equation}
The next proposition establishes that $\mathbb{X}_0$ is forward invariant for system $\mathcal{H}_I$ for any given $I\in\Rlp$, as required by Definition \ref{def:stability-hybrid-limit-cycle}\emph{(i)}. 

\begin{prop}\label{prop:never-at-full-rest} Given any $I\in\Rlp$, consider system $\mathcal{H_I}$ in (\ref{eq:hybrid-model-compact}). Set $\mathbb{X}_0$ in (\ref{eq:X0}) is forward invariant, i.e., $\rge(x)\subset\mathbb{X}_0$.    
\end{prop}

\noindent\textbf{Proof:} Let $I\in\Rlp$ and consider system $\mathcal{H}_I$.  First, we note that $\mathcal{C}\cup\mathcal{D}$ is forward invariant as $\mathcal{G}(\mathcal{D})\subset\mathcal{C}\cup\mathcal{D}$. Now, let $x=(q_1,q_2,\sigma)$ be a solution to $\mathcal{H}_I$ initialized in $\mathbb{X}_0$. We need to show that for  any $(t,j)\in\dom x$, $(q_1(t,j),q_{2}(t,j))\neq(0,0)$. We proceed by contradiction and consider the first hybrid time $(t,j)\in\dom x$ such that $x(t,j)\notin\mathbb{X}_0$. We have that $(t,j-1)\notin\dom x$ as $\mathcal{G}(\mathcal{D})\not\subset\{0\}\times\{0\}\times\{-1,1\}$. As a consequence, there exists $\varepsilon>0$ such that $(t-\varepsilon,j)\in\dom x$. The flow map $f$ is linear and  time-invariant, it thus generates unique solutions on flows. As $f(0,0,\{-1,1\})=(0,0,\{-1,1\})$, this implies that  $(q_1(t-\varepsilon,j),q_2(t-\varepsilon,j))=(0,0)$, i.e., $x(t-\varepsilon,j)\notin\mathbb{X}_0$. This contradicts the fact that $(t,j)$ is the first hybrid time in $\dom x$ such that $x(t,j)\notin\mathbb{X}_0$. Consequently, set $\mathbb{X}_0$ is forward invariant.  \hfill $\blacksquare$

In addition to Proposition \ref{prop:never-at-full-rest}, for any initial condition in $\mathbb{X}_0$, there exists a unique solution to $\mathcal{H}_I$, as formalized next.

\begin{lem}\label{lem:unique-solutions-in-X0} Given any $I\in\Rlp$, consider system $\mathcal{H}_I$ in (\ref{eq:hybrid-model-compact}). For any $x_0\in\mathbb{X}_0$, there exists a unique maximal solution initialized at $x_0$. 
\end{lem}

\noindent\textbf{Proof:} Let $I\in\Rlp$. Since $\mathbb{X}_0$ is forward invariant for system $\mathcal{H}_I$ by Proposition \ref{prop:never-at-full-rest}, we can consider the hybrid system denoted $\widetilde{\mathcal{H}}_I$ with flow map $f$, jump map $G$, flow set $\mathcal{C}\cap\mathbb{X}_0$ and jump set $\mathcal{D}\cap\mathbb{X}_0$. As $f$  is linear and $G$ is single-valued on $\mathbb{X}_0$, the only possibility for system $\widetilde{\mathcal{H}}_I$ to generate non-unique solutions for some initial states is if there exists $x_0=(q_{1,0},q_{2,0},\sigma_0)\in\mathcal{C}\cap\mathcal{D}\cap \mathbb{X}_0$ such that $f(x_0)\in\mathcal{T}_{\mathcal{C}\cap\mathbb{X}_0}(x_0)$. Suppose this is the case, then $\sigma_0 q_{1,0}=0$ and $\sigma_0 q_{2,0}=0$, which is equivalent to $q_{1,0}=0$ and $q_{2,0}=0$ as $\sigma_0\neq 0$, hence  $x_0\notin\mathbb{X}_0$: we have obtained a contradiction. This concludes the proof. \hfill $\blacksquare$

\subsection{Solution expression}\label{subsect:explicitSolution}

The flow map of system \eqref{eq:hybrid-model-compact} is  linear and time-invariant system, moreover the jump map generates increases or decreases of $q_2$ depending on its sign  of fixed amplitude. Given these two properties, we can provide expressions of the solution to system \eqref{eq:hybrid-model-compact} for any given $I$, which will be useful for determining the existence and stability of a hybrid limit cycle in the sequel. Its proof is omitted for space reasons, it is obtained by direct computations in view of (\ref{eq:hybrid-model-compact}). 

\begin{prop}\label{Prop:explicitSolution}
Given any $I\in\Rlp$, consider system $\mathcal{H_I}$ in (\ref{eq:hybrid-model-compact}). Let $x=(q_1,q_2,\sigma)$ be a solution initialized in $\mathbb{X}_0$ as defined in (\ref{eq:X0}) and $(t,j)\in\dom x$. For all $s \in [0,t_1]$,
   \begin{equation}
	\begin{array}{rlllll} 
		q_1(s,0)& = & q_1(0,0)e^{as}\cos{(bs)} +  c_0(x(0,0)) e^{as} \sin{(bs)} \\
		q_2(s,0)& = & q_2(0,0) e^{as}\cos{(bs)} \\
              && +(c_0(x(0,0))a -  q_1(0,0)b) e^{as}\sin{(bs)}\\
        \sigma(s,0) & \in & \SGN(q_1(0,0)), 
	\end{array}
	\label{eq:EqFlowSolutionProposition_i0}
\end{equation}
and, for each $i \in \{1,2, \dots, j\}$, and all $s \in [t_i, t_{i+1}]$, 
\begin{equation}
	\begin{array}{rlllll} 
		q_1(s,i)& = & c(q_2(t_i,i-1)) e^{a(s-t_i)} \sin{(b(s-t_i))} \\
		q_2(s,i)& = &  c(q_2(t_i,i-1)) e^{a(s-t_i)}[a \sin{(b(s-t_i))} \\
         & &+ b \cos{(b(s-t_i))}] \\
        \sigma(s,i) & = & \left\{\begin{array}{lllll}(-1)^{i}\sigma(0,0) \text{ if }  (q_1(0,0),q_2(0,0))\neq 0 \\ 
        (-1)^{i-1}\sign(q_2(0,1))   \text{ otherwise}
        \end{array}\right.
	\end{array}
\label{eq:EqFlowSolutionProposition}
\end{equation}
with  $0 = t_0 \leq t_1 \leq \dots \leq t_{j+1} = t$ satisfy $\dom x \cap ([0,t]\times \{0,1,\dots,j\}) = \bigcup_{i=0}^j [t_i, t_{i+1}] \times \{i\}$, $a:= -\frac{\alpha}{2} \in (-1,0)$, $b:= \frac{\sqrt{4-\alpha^2}}{2} \in (0,1)$, $c_0(x(0,0))=\frac{q_2(0,0) -q_1(0,0) a }{b}\in\R$ and $c(z):= \frac{z + I\sign(z)}{b} \in \R\backslash\{0\}$ for any $z\in\R$. 
\end{prop}

Proposition \ref{Prop:explicitSolution} provides  expressions for the solutions to system $\mathcal{H}_I$ in \eqref{eq:hybrid-model-compact} for any $I\in\Rlp$. This result  allows concluding that, after a first jump, the amount of continuous time between two consecutive jumps  is fixed and the same for any solution to system (\ref{eq:hybrid-model-compact}) initialized in $\mathbb{X}_0$. 

\begin{lem}\label{lem:periodic-jumps} Given any $I\in\Rlp$, consider system $\mathcal{H}_I$ in (\ref{eq:hybrid-model-compact}). For any solution $x$ initialized in $\mathbb{X}_0$ and any $(t,j),(t',j)\in\dom x$ such that  $j\geq 1$ and $(t,j-1),(t',j+1)\in\dom x$, it holds that 
$t' - t  =  \frac{\pi}{b}$. 
\end{lem}

\noindent\textbf{Proof:} Let $I\in\Rlp$, $x$ be a solution to $\mathcal{H_I}$ in \eqref{eq:hybrid-model-compact} initialized in $\mathbb{X}_0$ and $(t,j),(t',j)\in\dom x$ such that  $j\geq 1$ and $(t,j-1),(t',j+1)\in\dom x$. By definition of the jump set $\mathcal{D}$ in (\ref{eq:def-C-D}) and as $\rge(x)\subset\mathbb{X}_0$,
	$t' = \inf\{s \geq t:\sigma(s,j)q_1(s,j)=0  \text{ and } \sigma(s,j) q_2(s,j) \leq 0\}$. 
We note that the minimal continuous time $s'\geq t$ such that $q_1(s',j)=0$ is $s'=t+\frac{\pi}{b}$ in view of (\ref{eq:EqFlowSolutionProposition}). Moreover, $q_2(s',j)=-c(q_2(t,j-1))e^{a\frac{\pi}{b}}b$ and $\sigma(s',j)=\sigma(t,j)=\sign(q_2(t,j-1))=\sign(q_2(t,j)))=\sign(c(q_2(t,j-1))b)$. Consequently, $\sigma(s',j)q_2(s',j)\leq 0$, which implies that $s'=t'$:  we have obtained the desired result. \hfill $\blacksquare$

\subsection{Existence and uniqueness of hybrid limit cycle}\label{subsect:existence-limit-cycle}

In the next theorem we prove the existence of a periodic solution for system \eqref{eq:hybrid-model-compact}  according to Definition \ref{def:periodic-solution-hybrid-limit-cycle}. 

\begin{thm} \label{Thm:ExistenceFlowPeriodicSolution}
    Given any $I\in\Rlp$, system $\mathcal{H}_I$ in \eqref{eq:hybrid-model-compact} generates a $(2\frac{\pi}{b},2)$-periodic solution, which  defines a compact hybrid limit cycle denoted $\mathcal{O}_I$. %
\end{thm}

\noindent\textbf{Proof:} Let $x^\star=(q_1^\star,q_2^\star,\sigma^\star)$ be the maximal solution to $\mathcal{H}_I$ initialized at $(0,\mu^{\star}(I),1)$ with $\mu^{\star}(I):=I\frac{e^{a\frac{\pi}{b}}}{e^{a\frac{\pi}{b}}-1}\in\R_{<0}$ as $a<0$ and $b>0$ by Proposition \ref{Prop:explicitSolution}. We are going to prove that $x^{\star}$ is a $(2\frac{\pi}{b},2)$-periodic solution. We can already mention that Definition \ref{def:periodic-solution-hybrid-limit-cycle}\emph{(i)} holds with $T^{\star}=2\frac{\pi}{b}$ and $N^{\star}=2$. 

On the other hand,  $x^\star(0,0)\in\mathcal{D}\backslash\mathcal{C}$ as $\sigma^{\star}(0,0)q_1^{\star}(0,0)=0$ and $\sigma^{\star}(0,0)q_{2}^{\star}(0,0)= \mu^{\star}(I)<0$. Hence $(0,1)\in\dom x^{\star}$. As $x^\star(0,0)\in\mathbb{X}_0$, Lemma \ref{lem:periodic-jumps} applies and $(\frac{\pi}{b},1),(\frac{\pi}{b},2)\in\dom x^{\star}$. We deduce, still from Lemma \ref{lem:periodic-jumps}, that $\dom x^{\star}=\bigcup_{j=0}^{\infty}([j\frac{\pi}{b},(j+1)\frac{\pi}{b}]\times\{j+1\})\cup (\{0\}\times\{0\})$. Consequently, for any $(t,j)\in\dom x^{\star}$, $(t+2\frac{\pi}{b},j+2)\in\dom x^{\star}$. We have proved that Definition \ref{def:periodic-solution-hybrid-limit-cycle}\emph{(ii)} holds. 

By Proposition \ref{Prop:explicitSolution}, $x^{\star}(\frac{\pi}{b},1)=(0,-e^{a\frac{\pi}{b}}(\mu^{\star}(I)-I),-1)$. By definition of $\mu^{\star}(I)$,
\begin{equation}
\begin{array}{rllll}
q_2^{\star}(\frac{\pi}{b},1) & = & -e^{a\frac{\pi}{b}}(\mu^{\star}(I)-I) \\
& = & -e^{a\frac{\pi}{b}}(I\frac{e^{a\frac{\pi}{b}}}{e^{a\frac{\pi}{b}}-1}-I)\\
& = & -I\frac{e^{a\frac{\pi}{b}}}{e^{a\frac{\pi}{b}}-1} = -q_2^{\star}(0,0).
\end{array}
\end{equation}
We derive by iteration that for any $j\in\Zo$, recalling that $(j\frac{\pi}{b},j)\in\dom x^{\star}$ by Lemma \ref{lem:periodic-jumps}, 
\begin{equation}\label{eq:proof-existence-periodic-at-jumps}
\begin{array}{rllll}
q_2^{\star}(j\frac{\pi}{b},j) & = & (-1)^{j}q_2^{\star}(0,0).
\end{array}
\end{equation}
Let $(t,j)\in\dom x^{\star}$. By Proposition \ref{Prop:explicitSolution}, noting that $t_{j+2}=(j+1)\frac{\pi}{b}$, and by (\ref{eq:proof-existence-periodic-at-jumps}),
\begin{equation}
	\begin{array}{rlllll} 
		q_1^{\star}(t+2\frac{\pi}{b},j+2)& = & c(q_2(t_{j+2},j+1)) e^{a(t+2\frac{\pi}{b}-t_{j+2})} \\
        & & \times \sin{(b(t+2\frac{\pi}{b}-t_{j+2}))}\\
        & = & (-1)^{j+2}q_{2}^{\star}(0,0)b^{-1} e^{a(t-(j-1)\frac{\pi}{b})} \\
        & & \times \sin{(b(t-(j-1)\frac{\pi}{b}))}\\
        & = & (-1)^{j}q_{2}^{\star}(0,0)b^{-1} e^{a(t-(j-1)\frac{\pi}{b})} \\
        & & \times \sin{(b(t-(j-1)\frac{\pi}{b}))}\\
        & = & q_1^{\star}(t,j).
\end{array}
\end{equation}
We similarly derive that $q_2^{\star}(t+2\frac{\pi}{b},j+2)=q_2^{\star}(t,j)$ and $\sigma^{\star}(t+2\frac{\pi}{b},j+2)=\sigma^{\star}(t,j)$. Hence Definition \ref{def:periodic-solution-hybrid-limit-cycle}\emph{(iii)} holds.

Finally, we derive from the fact that $x^{\star}$ is $(2\frac{\pi}{b},2)$-periodic and its expression coming from Proposition \ref{Prop:explicitSolution} that $\rge(x^{\star})$ is bounded and closed, i.e., it is compact. This allows concluding that  $\mathcal{O}_I=\rge(x^\star)$ is a compact hybrid limit cycle for system  $\mathcal{H}_I$.  \hfill $\blacksquare$

Not only system $\mathcal{H}_I$ generates $(2\frac{\pi}{b},2)$-periodic solutions for any given $I\in\Rlp$ as established in Theorem \ref{Thm:ExistenceFlowPeriodicSolution}, but all such periodic solutions that are initialized in $\mathbb{X}_0$ have the same range and  define the same, unique hybrid limit cycle $\mathcal{O}_I$. 

\begin{thm}\label{thm:unique-hybrid-limit-cycle} Given any $I\in\Rlp$, consider system $\mathcal{H_I}$. Set $\mathcal{O}_I$ in Theorem \ref{Thm:ExistenceFlowPeriodicSolution} is the unique hybrid limit cycle  in $\mathbb{X}_0$. 
\end{thm}

\noindent\textbf{Proof:} Let $I\in\Rlp$. We start by showing that for any $(2\frac{\pi}{b},2)$-periodic solutions $x$, $x'$ initialized in $\mathbb{X}_0$, $\text{rge}(x)=\text{rge}(x')$. 

Let $x=(q_1,q_2,\sigma)$ be a $(2\frac{\pi}{b},2)$-periodic solution to system $\mathcal{H}_I$ initialized in $\mathbb{X}_0$. Let $t_1,t_2\in\Rlo$ be such that $(t_1,0),(t_1,1),(t_2,1),(t_2,2)\in\dom x$. Since $x$ is $(2\frac{\pi}{b},2)$-periodic, we derive, similarly to the proof of Theorem \ref{Thm:ExistenceFlowPeriodicSolution} that $q_2(t_2,1)=-q_2(t_1,0)$. By Proposition \ref{Prop:explicitSolution} and Lemma \ref{lem:periodic-jumps}, this is equivalent to
\begin{equation}
\begin{array}{rllll}
-[q_2(t_1,0)+I\sign(q_2(t_1,0))]e^{a\frac{\pi}{b}} = - q_2(t_1,0)
\end{array}\label{eq:proof-thm-existence-limit-cycle}
\end{equation}
If $q_2(t_1,0)<0$, then (\ref{eq:proof-thm-existence-limit-cycle}) is equivalent to $q_2(t_1,0)=I\frac{e^{a\frac{\pi}{b}}}{e^{a\frac{\pi}{b}}-1}=\mu^{\star}(I)$ with $\mu^{\star}(I)$ defined in the proof of Proposition~\ref{Thm:ExistenceFlowPeriodicSolution}. In this case, we have $x(t_1,0)=(0,\mu^{\star}(I),1)$. This implies that $x(t_1,0)=x^{\star}(0,0)$ where $x^{\star}$ is the $(2\frac{\pi}{b},2)$-periodic solution constructed in the proof of Proposition~\ref{Thm:ExistenceFlowPeriodicSolution}. Consequently, by Lemma \ref{lem:unique-solutions-in-X0}, we derive that $\rge(x)=\rge(x^{\star})$. If $q_2(t_1,0)>0$, we can derive that $q_2(t_2,1)<0$  and $x(t_2,1)=(0,\mu^{\star}(I),1)$ and we similarly conclude that $\rge(x)=\rge(x^{\star})$. Hence,  for any $(2\frac{\pi}{b},2)$-periodic solutions $x$, $x'$ initialized in $\mathbb{X}_0$, $\text{rge}(x)=\text{rge}(x')=\rge(x^{\star})$. 

We similarly derive that for any $(2k\frac{\pi}{b},2k)$-periodic solutions $x$, $x'$ initialized in $\mathbb{X}_0$ with $k\in\Zp$, $\text{rge}(x)=\text{rge}(x')$. 

Consider a $(\widetilde T,\widetilde N)$-periodic solution denoted $\tilde x=(\tilde q_1,\tilde q_2,\tilde\sigma)$ initialized in $\mathbb{X}_0$. As $\tilde q_2$ and $\tilde \sigma$ both change sign at each jump and $\tilde x$ jumps every $\frac{\pi}{b}$ continuous time (after the first jump) by Lemma \ref{lem:periodic-jumps}, necessarily $\widetilde N\geq 2$ and $\widetilde N$ is even. This property of $\widetilde N$ together with the fact, again, that two consecutive jumps are spaced by exactly $\frac{\pi}{b}$ units of continuous time (after the first jump) by Lemma \ref{lem:periodic-jumps} implies that there exists $k\in\Zp$ such that $\widetilde T=k2\frac{\pi}{b}$ and $\widetilde N=2k$. As $x^\star$ is also $(2k\frac{\pi}{b},2k)$-periodic, we derive that $\text{rge}(\widetilde x)=\rge(x^\star)=\mathcal{O}_I$, which concludes the proof. \hfill $\blacksquare$

\hide{
\textcolor{red}{add few lines on comment on the meaning of this theorem... and that $(0,-\bar q_2^\star)$ is the fixed point of the hybrid Poincaré map.. maybe in a definition? Not sure, I need to think about how to write this nicely.}}

\hide{\textcolor{red}{Define $P$ as the hybrid Poincaré map for \eqref{eq:hybrid-model-compact} with fixed point $\bar x^\star:= (0,-\bar q_2^\star)$.. we need to use this in the proof below.. I need to find a way to define it.. maybe inspired by \cite[equation (8)]{lou-et-al-tac18(hybrid-limit-cycle)}??}}

\subsection{Stability properties of the hybrid limit cycle}\label{subsec:StabilityHybridLimitCycle}

In the next theorem we prove the existence and a stability property for a hybrid limit cycle of system \eqref{eq:hybrid-model-compact}.  

\begin{thm}\label{Thm:StabilityLimitCycle} Given any $I\in\Rlp$, consider system $\mathcal{H}_I$. The hybrid limit cycle $\mathcal{O}_I$ in Theorem \ref{Thm:ExistenceFlowPeriodicSolution} is  uniformly exponentially stable on $\mathbb{X}_0$ in (\ref{eq:X0}). 
\end{thm}

\noindent\textbf{Proof:}  Let $I\in\Rlp$. Given any $x_0=(q_{1,0},q_{2,0},\sigma_{0})\in\mathbb{X}_0$, we introduce by $\tau$ the time-to-impact function \cite{lou-et-al-tac18(hybrid-limit-cycle),grizzle2002asymptotically} 
\begin{equation}
\begin{array}{rlll}
\tau(x_0):=\inf\{s\geq 0\,:\,\phi_1(x_0,s)=0,\,\phi_2(x_0,s)< 0\},
\end{array}
\end{equation}
with $\phi_1(x_0,s):=q_{1,0}e^{as}\cos{(bs)} +  c_0(x_0) e^{as} \sin{(bs)}$ and $\phi_2(x_0,s):=\sigma_0 (q_{2,0} e^{as}\cos{(bs)}  +(c_0(x_0)a -  q_{1,0}b) e^{as}\sin{(bs)})$ from Proposition \ref{Prop:explicitSolution}. Given any initial condition $x_0\in\mathbb{X}_0$, $\tau(x_0)$ is  the first continuous time at which a jump occurs for the solution to $\mathcal{H}_I$ initialized at $x_0$. We derive from \cite[Lemma 3]{grizzle2002asymptotically} that for any $\sigma_0\in\{-1,1\}$, the mapping $\tilde\tau:q_0\mapsto\tau(q_0,\sigma_0)$ is continuous on $\R^{2}\backslash\{(0,0)\}$. This property will be important later in the proof.

Let $x=(q_1,q_2,\sigma)$ be any maximal solution to $\mathcal{H}_I$ initialized in $\mathbb{X}_0$ and we suppose that $\sigma(0,0)=1$ without loss of generality as the proof similarly follows when $\sigma(0,0)=-1$. Let $\bar x=(\bar q_1,\bar q_2,\bar\sigma)$ be the maximal solution to $\mathcal{H}_I$ initialized at $(M(\tau(x(0,0))(0,
\mu^\star(I)e^{-a\tau(x(0,0))}),1)$ where for any $s\in\Rlo$,
\begin{equation}
\begin{array}{rllll}
M(s):=\left[\begin{smallmatrix} \cos(bs)+\frac{a}{b}\sin(bs) & -\sin(bs)\\ (\frac{a^{2}}{b}+b)\sin(bs)& b\cos(bs)-a\sin(bs)\end{smallmatrix}\right].
\end{array}\label{eq:M(s)}
\end{equation}
We  have that $\bar x(\tau(x(0,0)),0)=(0,\mu^{\star}(I),1)$ by direct computations using Proposition \ref{Prop:explicitSolution}. Therefore, we deduce like in the proof of Theorem \ref{thm:unique-hybrid-limit-cycle} that  $\bar x$ is $(2\frac{\pi}{b},2)$-periodic and $\dom \bar x=\dom x$.

Let $e:=x-\bar x$. For all $j\in\Zo$ and almost all $t\in[t_j,t_{j+1}]$ with the notation of Proposition \ref{Prop:explicitSolution}, $\dot e(t,j) = A e(t,j)$ with $A=\left[\begin{smallmatrix}0 & 1 \\ -1 & -\alpha\end{smallmatrix}\right]$. Moreover, for any $j\in\Zo$,  $e(t_{j+1},j+1)=e(t_{j+1},j)$. As $A$ is Hurwitz, we derive  that there exist  $d_1\geq 1$ and $d_2>0$ independent of $x$ and $\bar x$ such that for all $(t,j)\in\dom x$, $|e(t,j)|\leq d_1 e^{-d_2 t}|e(0,0)|$. Let $(t,j)\in\dom x$. By Lemma \ref{lem:periodic-jumps}, we have that $t\geq (j-1)\frac{\pi}{b}$. Hence $\frac{t}{2}\geq \frac{1}{2}(j-1)\frac{\pi}{b}$ from which we derive
\begin{equation}
\begin{array}{rllll}
|e(t,j)| & \leq &  d_1 e^{-\frac{d_2}{2}(t+(j-1)\frac{\pi}{b})}|e(0,0)|\\
& \leq & d_1 e^{\frac{d_2 \pi}{2b}} e^{-\frac{d_2}{2}\min(1,\frac{\pi}{b})(t+j)}|e(0,0)| \\
& = & \gamma_1 e^{-\gamma_2(t+j)}|e(0,0)|,
\end{array}
\end{equation}
with $\gamma_1:=d_1 e^{\frac{d_2 \pi}{2b}}\geq 1$ and $\gamma_2:=\frac{d_2}{2}\min(1,\frac{\pi}{b})=\frac{d_2}{2}>0$ as $b \in (0,1)$.

Now consider $|x(t,j)|_{\mathcal{O}_I}$ with $\mathcal{O}_I$ as in Theorem \ref{Thm:ExistenceFlowPeriodicSolution}. Since $\rge(\bar x)=\mathcal{O}_I$ by Theorem \ref{thm:unique-hybrid-limit-cycle},
\begin{equation}
\begin{array}{rlll}
|x(t,j)|_{\mathcal{O}_I} & \leq & |e(t,j)| \\
& \leq & \gamma_1 e^{-\gamma_2(t+j)}|e(0,0)|.
\end{array}\label{eq:proof-thm-stability-exp-KL}
\end{equation}
It holds that $|e(0,0)|=|(q_1(0,0),q_2(0,0))-M(\tau(x(0,0)))(0,
\mu^\star(I)e^{-a\tau(x(0,0))})|$. We introduce the mapping 
\begin{equation}
\psi_1\!:\!z\!=\!(z_1,z_2)\mapsto |(z_1,z_2)-M(\tau(z,1))(0,\mu^{\star}(I)e^{-a\tau(z,1)})|
\end{equation}
defined over $\mathbb{X}_0$. The mapping $\psi_1$ is  continuous as so are $\tau$  and $M$ in view of (\ref{eq:M(s)}). Moreover  $\psi_1(z)=0$
if and only if $z\in\mathcal{O}_I$. Indeed, if $z\in\mathcal{O}_1$, $M(\tau(z,1))^{-1}z=(0,\mu^{\star}(I)e^{-a\tau(z,1)})$, noting that $M(s)$ is invertible for any $s\in\Rlp$ as its determinant is equal to $1$. This implies that a solution initialized at $(z,1)$ reaches $(0,\mu^{\star}(I),1)$ in $\tau(z,1)$ units of continuous time, which implies that this solution is $(2\frac{\pi}{b},2)$-periodic as advocated in the proof of Theorem \ref{thm:unique-hybrid-limit-cycle}. Consequently $z\in\mathcal{O}_I$ still by Theorem \ref{thm:unique-hybrid-limit-cycle}. On the other hand, if $z\notin\mathcal{O}_I$, this means that $z\neq M(\tau(z,1))(0,\mu^{\star}(I)e^{-a\tau(z,1)})$ and thus $\psi_1(z)>0$. Lastly, $\psi_1$ grows unbounded with $(z_1,z_2)$ as $\tau$ takes value in $[0,\frac{\pi}{b}]$ by Lemma \ref{lem:periodic-jumps}. As a consequence, noting that $\mathcal{O}_I$ is compact by Theorem \ref{Thm:ExistenceFlowPeriodicSolution}, we deduce from these properties of $\psi_1$ and \cite[p.54]{goebel2012hybrid} that there exists $\alpha_1\in\Kinf$ such that for any $z\in\mathcal{O}_I$, $\psi_1(z)\leq\alpha_1(|z|_{\mathcal{O}_I})$. Therefore, by (\ref{eq:proof-thm-stability-exp-KL}), for any $(t,j)\in\dom x$,
\begin{equation}
\begin{array}{rlll}
|x(t,j)|_{\mathcal{O}_I} & \leq &  \gamma_1 e^{-\gamma_2(t+j)}\alpha_1(|x(0,0)|_{\mathcal{O}_I}).
\end{array}\label{eq:proof-stability-sigma1}
\end{equation}
We have proved that there exist $\gamma_1\geq 1$, $\gamma_2>0$ and $\alpha_1\in\Kinf$ such that for any solution $x=(q_1,q_2,\sigma)$ to system $\mathcal{H}_I$ initialized in $\mathbb{X}_0$ with $\sigma(0,0)=1$, (\ref{eq:proof-stability-sigma1}) holds. We similarly derive that there exist $\alpha_{-1}\in\Kinf$ such that such that for any solution $x=(q_1,q_2,\sigma)$ to system $\mathcal{H}_I$ initialized in $\mathbb{X}_0$ with $\sigma(0,0)=-1$, $|x(t,j)|_{\mathcal{O}_I} \leq \gamma_1 e^{-\gamma_2(t+j)}\alpha_{-1}(|x(0,0)|_{\mathcal{O}_I})$ for all $(t,j)\in\dom x$. Consequently, Definition \ref{def:stability-hybrid-limit-cycle}\emph{(ii)} holds with $\alpha=\gamma_1\max\{\alpha_1,\alpha_{-1}\}\in\Kinf$ and $\gamma=
\gamma_2>0$. \hfill $\blacksquare$

\begin{rem}[Robustness] The stability property established in Theorem \ref{Thm:StabilityLimitCycle} exhibits intrinsic robustness properties due to the fact that $\mathcal{O}_I$ is compact and that system $\mathcal{H}_I$ satisfies the hybrid basic conditions, see, e.g. \cite[Chapter 7]{goebel2012hybrid}. For instance, we have from \cite[Proposition 2.4]{cai2009characterizations} that system $\mathcal{H}_I$ is locally input-to-state stable with respect to any  disturbances entering continuously in the flow and the jump maps for any $I\in\Rlp$. These disturbances may model for instance the mismatch between system (\ref{eq:sys-pendulum-q1-q2-flow}) and its linearization in (\ref{eq:sys-pendulum-q1-q2-flow_linearized}).
\end{rem}

\section{Numerical illustration}\label{sect:numerical-illustration}
\begin{figure}[t!]
	\centering
	\includegraphics[scale=0.45]{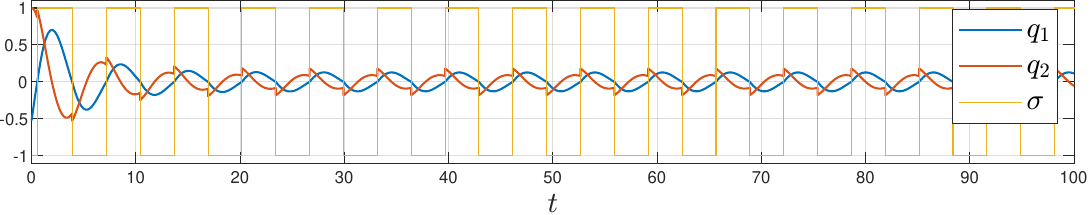}
	\caption{Angular position (red), velocity (blue) and $\sigma$ (yellow) with nonlinear dynamics and initial condition $(\frac{\pi}{3},2,-1)$.\vspace{-0cm}}
	\label{fig:simulation}
\end{figure}

We consider the nonlinear dimensionless pendulum dynamics in \eqref{eq:pendulum}, with $\alpha = 0.5$ and spiking input $u$ defined in \eqref{eq:spikingInput} with $I = 0.1$. In particular, we consider system \eqref{eq:hybrid-model-compact} where the nonlinear dynamics \eqref{eq:sys-pendulum-q1-q2-flow} is considered in the flow map \eqref{eq:def-f-G}, instead of \eqref{eq:sys-pendulum-q1-q2-flow_linearized}. 
%
The simulation results obtained with initial conditions $q_1(0,0) = \frac{\pi}{3}$, $q_2(0,0) = 2$ and $\sigma(0,0) = -1$ are depicted in Fig. \ref{fig:simulation}. 
\begin{figure}[t!]
	\centering
	\includegraphics[scale=0.4]{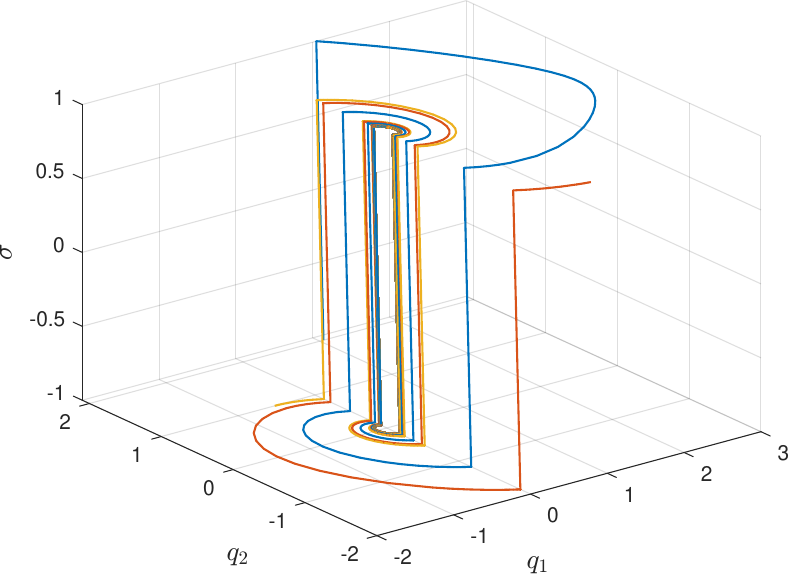}
	\caption{Hybrid limit cycle with nonlinear dynamics with different initial conditions.\vspace{-0cm}}
	\label{fig:simulation3}
\end{figure}
Fig. \ref{fig:simulation} shows that the considered solutions to the closed-loop pendulum system converges to a hybrid limit cycle, as guaranteed by Theorem \ref{Thm:StabilityLimitCycle}. Moreover, we have run three  simulations considering three different initial conditions. The results are depicted in Fig. \ref{fig:simulation3}, where in the blue plot we used $(\frac{\pi}{3}, 2,  -1)$, in the red one $( \frac{\pi}{4},-2, 1)$ and in the yellow plot we considered $(-\frac{\pi}{6},1, -1)$. Fig. \ref{fig:simulation3} shows that, for all the three considered initial conditions, the solutions converge to the same hybrid limit cycle consistently with Theorem \ref{Thm:StabilityLimitCycle}.

\section{Discussion on pulse amplitude adaptation}\label{sect:discussion}


A next relevant step is to  control the maximal amplitude of the hybrid limit cycle by introducing an adaptation loop on the control parameter $I \in \Rlp$. This is done in \cite{schmetterling2024neuromorphic} as well as in the companion paper \cite{medvedeva-et-al-cdc25} where the neurons activation is being tuned and analytical guarantees are provided.  In this section, we  briefly elaborate on this research direction by adding an adaptation loop for $I$ to system $\mathcal{H}_I$ in (\ref{eq:hybrid-model-compact}), together with associated numerical simulations showing interesting and promising results. A formal analysis of this augmented hybrid system is left for future work. 

\subsection{Adaptation rule}\label{subsect:adaptationLoop}
To control the amplitude of the pendulum oscillations we introduce in this section a ``spiky'' adaptation loop on the parameter $I \in \Rlp$ in \eqref{eq:spikingInput}. In view of the pendulum dynamics \eqref{eq:sys-pendulum-q1-q2-flow} and the explicit solution in Proposition \ref{Prop:explicitSolution}, the maximal amplitude of the oscillations is obtained whenever the angular velocity $q_2$ changes sign, which can be measured using neuromorphic devices \cite{schmetterling2024neuromorphic,medvedeva-et-al-cdc25}. 
Based on the value of the actual maximal amplitude and the desired one, the parameter $I \in \Rlp$ is adapted. In particular, when $q_2$ is equal to $0$, a spiking adaptive action with amplitude $\varepsilon \in \Rlp$ on the control parameter $I$ is generated, whose sign depends on the sign of the difference between the measured amplitude of the oscillations, given by $\sigma_1 q_1 \in (0,\pi]$ where $\sigma_1$ denotes variable $\sigma$ in the previous sections, and the desired one, denoted by $q_1^\star \in (0,\pi)$, i.e., in terms of the hybrid systems notation in Section \ref{Notation},
\begin{equation}
		I^+ \in \{I -\varepsilon w\,:\,w\in\SGN(\sigma q_1 - q_1^\star)\} =: G_I(q_1,\sigma,I),
        \label{eq:TorqueJumpAdaptation}
	\end{equation}
and, on flows,
\begin{equation}
     \dot I = 0. 
\end{equation}
This learning (adaptation) loop is ``spiky'' in nature and thus sparse. 
To add this adaptation to the hybrid system $\mathcal{H}_I$ in a robust way (especially the jump \eqref{eq:TorqueJumpAdaptation} related to $q_2 = 0$), 
similarly to \eqref{eq:sigmaFlow} and \eqref{eq:sigmaJump} we introduce an auxiliary logic variable $\sigma_2 \in \{-1,1\}$ to robustify the detection of the condition $q_2 =0$. In particular, on flows,
\begin{equation}
	\dot\sigma_2 = 0,
	\label{eq:sigma2Flow}
\end{equation}
and at jumps, when $q_2 = 0$, 
\begin{equation}
	\sigma_2^+ = -\sigma_2. 
	\label{eq:sigma2Jump}
\end{equation}
When $q_2$ changes sign and thus the parameter $I \in \Rlp$ is updated according to \eqref{eq:TorqueJumpAdaptation}, the pendulum angular position, velocity and the auxiliary signal $\sigma$, which in the augmented hybrid model we call $\sigma_1$, are not modified, i.e., $q_1^+ = q_1$, $q_2^+ = q_2$ and $\sigma_1^+ = \sigma_1$. 

Using \eqref{eq:sys-pendulum-q1-q2-flow}, \eqref{eq:sys-pendulum-q1-q2-jump}, \eqref{eq:sigmaFlow}, \eqref{eq:sigmaJump} and \eqref{eq:TorqueJumpAdaptation}-\eqref{eq:sigma2Jump}, in the next section we define an augmented  hybrid model, where a jump corresponds to a spiking torque action, as in \eqref{eq:hybrid-model-compact}, or to an update of the parameter $I \in \Rlp$, according to \eqref{eq:TorqueJumpAdaptation}. 



\subsection{Augmented hybrid model}
Inspired by system $\mathcal{H}_I$ described in Section \ref{subsect:hybridModel} and using \eqref{eq:TorqueJumpAdaptation}-\eqref{eq:sigma2Jump}, we now define an augmented hybrid model, which includes the adaptation loop described in Section \ref{subsect:adaptationLoop}. Contrary to $\mathcal{H}_I$, in this section we consider the nonlinear pendulum dynamics during flow \eqref{eq:sys-pendulum-q1-q2-flow} and not the linearized one \eqref{eq:sys-pendulum-q1-q2-flow_linearized} as no analytical guarantees are provided.

Given a desired maximal angular amplitude $q_1^\star  \in (0,\pi)$, we denote the concatenated state variable 
\begin{equation}
	\begin{array}{rlllll} 
		\bar x & := & (q_1,q_2,\sigma_1, \sigma_2, I)\in \overline{\mathbb{X}}\subset\R^{n_{\bar x}}\\
		\overline{\mathbb{X}} & := & \R^2\times\{-1,1\}^2\times\R_{\geq 0},
	\end{array}
\end{equation}
with $n_{\bar x}:=5$. We propose the hybrid model
\begin{equation}\label{eq:hybrid-model-compact_learning}
	\begin{array}{rlllll} 
		\overline{\mathcal{H}}\,:\,\left\{\begin{array}{rlllll}\dot{\bar x} & = & \overline{f}(\bar x)  & & \bar x\in\overline{\mathcal{C}}\\
			\bar x^+ & \in & \overline G(\bar x)   & & \bar x\in\overline{\mathcal{D}},
		\end{array}\right.
	\end{array}
\end{equation}
with 
\begin{equation}\label{eq:def-C-D_learning}
	\left\{\begin{array}{rlllll} 
		\overline{\mathcal{C}} & := & \left\{\bar x\in\overline{\mathbb{X}}\,:\,\sigma_1 q_1\geq 0, \sigma_2 q_2\geq 0\right\} \\
		\overline{\mathcal{D}} & := & \overline{\mathcal{D}}_1 \cup \overline{\mathcal{D}}_2\\
		\overline{\mathcal{D}}_1 & := & \left\{\bar x\in\overline{\mathbb{X}}\,:\,\sigma_1 q_1= 0,\,\sigma_1 q_2\leq 0,\,\sigma_2 q_2\geq 0\right\} \\
		\overline{\mathcal{D}}_2 & := & \left\{\bar x\in\overline{\mathbb{X}}\,:\,\sigma_1 q_1\geq 0, \sigma_2 q_2=0,\,\right.\\
        & & \quad\quad\quad\quad\left.-a\sigma_2\sin(q_1)\leq 0\right\} 
	\end{array}\right.\nonumber
\end{equation}
and $\overline f(\bar x) := (q_2,- \sin{(q_1)} -\alpha q_2,0, 0, 0)$ for any  $\bar x\in\overline{\mathcal{C}}$, $\overline G(\bar x)  :=  \overline G_1(\bar x) \cup \overline G_2(\bar x)$ for any $ \bar x\in\overline{\mathcal{D}}$, 
\begin{equation}\label{eq:def-f-G_learning}
	\begin{array}{rlllll} 
		\overline G_1(\bar x) \hspace{-0.3cm}& := & \hspace{-0.3cm} \left\{\begin{array}{lllll}
			\hspace{-0.2cm}\{(q_1,q_2 + Iz,z, \sigma_2, I):z\in\SGN{(q_2)}\},  & \hspace{-0.2cm} \bar x\in \overline{\mathcal{D}}_1 \\
			\emptyset,  &\hspace{-0.2cm}  \bar x\notin \overline{\mathcal{D}}_1\\
		\end{array}\right.\\
        \overline G_2(\bar x) \hspace{-0.3cm}&:= & \hspace{-0.3cm}\left\{\begin{array}{lllll}
           \hspace{-0.2cm} \{(q_1,q_2,\sigma, -\sigma_2,I -\varepsilon w\,\!:\!\,w\in\SGN(\sigma q_1 - q_1^\star)\},\\ 
           \hspace{5.5cm}\bar x\in \overline{\mathcal{D}}_2 \\
			\emptyset,  \hspace{5.1cm}   \bar x\notin \overline{\mathcal{D}}_2. \\
		\end{array}\right.\nonumber
	\end{array}
\end{equation}

The next result follows similarly as Proposition \ref{prop:hybrid-model-basic-properties}, its proof is therefore omitted.

\begin{prop}\label{prop:hybrid-model-basic-properties_learning}
Given any $q_1^\star\in(0,\pi)$, the following hold for system $\overline{\mathcal{H}}$ in (\ref{eq:hybrid-model-compact_learning}).
\begin{enumerate}
\item[(i)] The hybrid basic conditions are satisfied.
\item[(ii)] Maximal solutions are complete.
\end{enumerate}
\end{prop}

\subsection{Simulation results}
\begin{figure}[t!]
	\centering
	\includegraphics[scale=0.35]{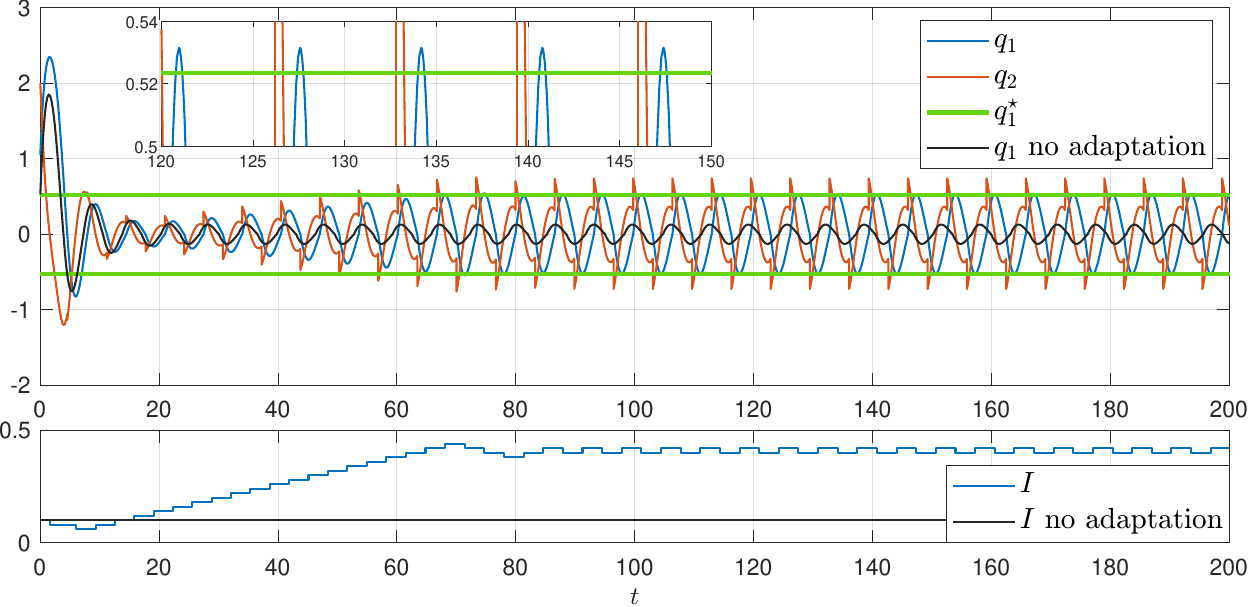}
	\caption{Angular position (blue) and velocity (red) with nonlinear dynamics and spiking adaptation loop to adapt the amplitude of the spikes $I$. Angular position (black) without adaptation (constant $I = 0.1$). 
}
	\label{fig:simulation2}
\end{figure}

We consider system \eqref{eq:hybrid-model-compact_learning} with $\alpha = 0.5$, $q_1^\star = \frac{\pi}{6}$ and $\varepsilon = 0.02$. The simulation results obtained for the pendulum angular position and velocity with initial conditions $q_1(0,0) = \frac{\pi}{3}$, $q_2(0,0) = 2$, $\sigma_1(0,0) = 1$, $\sigma_2(0,0) = 1$ and $I(0,0) = 0.1$  are shown in Fig. \ref{fig:simulation2}, together with the angular position obtained without adaptation mechanism, i.e., simulating system $\mathcal{H}_I$ with nonlinear flow dynamics from the same initial conditions, with $I \equiv 0.1$. Fig. \ref{fig:simulation2} also depicts
the adaptation of the parameter $I \in \Rlp$, which is linked to the amplitude of the spiking control actions inducing the limit cycle, 
 and is thus used to control the maximal amplitude of the oscillations.  
Fig. \ref{fig:simulation2} shows that including the spiking adaptation loop in \eqref{eq:TorqueJumpAdaptation}-\eqref{eq:sigma2Jump} allows generating a hybrid limit cycle with a desired maximal angular amplitude, and thus illustrates the potential of hybrid system models and tools for this purpose. Note that, in view of the fixed amplitude $\varepsilon \in \Rlp$ of the spiking adaptation mechanism in \eqref{eq:TorqueJumpAdaptation}, we expect a practical convergence property of the maximal amplitude of the oscillations $\sigma q_1$ to the desired one $q_1^\star$, 
as illustrated in the zoom in Fig.~\ref{fig:simulation2}. A formal analysis of system \eqref{eq:hybrid-model-compact_learning} is left for future work. 

\section{Conclusion}\label{sect:conclusion}

We have shown how hybrid system theoretic tools can be exploited to show the existence of a limit cycle and analyze its stability property for a pendulum controlled by a neuromorphic device. We relied on two main assumptions for this purpose: we have linearized the pendulum dynamics and abstracted the effect of bursting neurons, see \cite{schmetterling2024neuromorphic}, as Dirac impulses. We plan to relax these assumptions in the future by notably exploiting monotonocity properties of the effect of the bursting neurons activation on the angular velocity. We also plan to formalize the results hinted in Section \ref{sect:discussion} by exploiting hybrid singularly perturbed techniques (see, e.g., \cite{wang2012analysis,tanwani2024singularly}) thereby leading to a rigorous learning mechanism of the amplitude $I$, which is related to the length of activation of the bursting neurons in \cite{schmetterling2024neuromorphic}. 


\section*{Acknowledgements}

We would like to thank Alessio Franci, Fernando Casta\~nos, and Taisia Medvedeva for discussions on the complementary paper \cite{medvedeva-et-al-cdc25} and for providing their insights on the present study. Both papers are submitted to the same invited session on neuromorphic systems and control. We also thank Mattia Giaccagli and Pietro Lorenzetti for initial discussions, which have led to this paper.

\bibliography{IEEEabrv,bibliography}

\end{document}